\documentclass{article}[12pt]
\usepackage{cite}
\usepackage{graphicx}
\usepackage{epsfig}
\usepackage{amsfonts}
 \usepackage{amssymb}

\date{\today}

\newcommand{\insertplot}[5]{\begin{figure}
 \hfill\hbox to 0.05in{\vbox to #5in{\vfill
 \inputplot{#1}{#4}{#5}}\hfill}
 \hfill\vspace{-.1in}
 \caption{#2}\label{#3}
 \end{figure}}
 \newcommand{\inputplot}[3]{
 \special{ps: plotfile #1}
}
\newcounter{fig}

\newcommand{\ee}{\end{equation}}
\newcommand{\eea}{\end{eqnarray}}
\newcommand{\be}{\begin{equation}}
\newcommand{\bea}{\begin{eqnarray}}

\begin{document}

 \title{\bf {\LARGE
The scalarised Schwarzschild-NUT spacetime}}

\author{
{\large Yves Brihaye}$^{\dagger}$,
{\large Carlos Herdeiro}$^{\ddagger}$  
and {\large Eugen Radu}$^{\diamond}$  
\\ 
\\
$^{\dagger}${\small Physique-Math\'ematique, Universite de
Mons-Hainaut, Mons, Belgium}
\\
$^{\ddagger}${\small Centro de Astrof\'\i sica e Gravita\c c\~ao - CENTRA,} \\{\small Departamento de F\'\i sica,
Instituto Superior T\'ecnico - IST, Universidade de Lisboa - UL,} \\ {\small Avenida
Rovisco Pais 1, 1049-001, Portugal}
\\
$^{\diamond}${\small Departamento de F\'isica da Universidade de Aveiro and CIDMA,} \\ {\small Campus de Santiago, 3810-183 Aveiro, Portugal}
}
  
  \date{October 2018}

\maketitle

\begin{abstract}  
It has recently been suggested that vacuum  black holes of General Relativity (GR) can become spontaneously scalarised when appropriate non-minimal couplings to curvature invariants are considered. These models circumvent the standard black hole no scalar hair theorems of GR, allowing both the standard GR solutions and new scalarised ($a.k.a.$ hairy) solutions, which in some cases are thermodynamically preferred. Up to now, however, only (static and spherically symmetric) scalarised Schwarzschild solutions have been considered. It would be desirable to take into account the effect of rotation; however, the  higher curvature invariants introduce a considerable challenge in obtaining the corresponding scalarised rotating black holes.  As a toy model for rotation, we present here the scalarised generalisation of the Schwarzschild-NUT solution, taking either the Gauss-Bonnet (GB) or the Chern-Simons (CS) curvature invariant. The NUT charge $n$ endows spacetime with ``rotation", but the angular dependence of the corresponding scalarised solutions factorises, leading to a considerable technical simplification. For GB, but not for CS, scalarisation occurs for $n=0$. This basic difference leads to a distinct space of solutions in the CS case, in particular exhibiting a double branch structure. In the GB case, increasing the horizon area demands a stronger non-minimal coupling for scalarisation; in the CS case, due to the double branch structure, both this and the opposite trend are found. 
 We briefly comment also  on the scalarised Reissner-Nordstr\"om-NUT solutions.
\end{abstract}
 
  \newpage
 
%%%%%%%%%%%%%%%%%%%%%%%%%%%%%%%%%%%%%%%%%%%%%%%%%%%%%%%%%%%%%%%%%%%%%%%%%%%%%%%
\section{Introduction}
%%%%%%%%%%%%%%%%%%%%%%%%%%%%%%%%%%%%%%%%%%%%%%%%%%%%%%%%%%%%%%%%%%%%%%%%%%%%%%%
 It has long been known that violations of the strong equivalence principle, via the inclusion of non-minimal couplings, can lead to asymptotically flat black hole (BH) scalar ``hair" - see
~\cite{Bekenstein:1974sf,Bekenstein:1975ts,Bocharova:1970skc,Kanti:1995vq,Babichev:2013cya,Sotiriou:2013qea,Lee:2018zym} and~\cite{Herdeiro:2015waa,Sotiriou:2015pka,Volkov:2016ehx} for recent reviews. More recently~\cite{Doneva:2017bvd,Silva:2017uqg,Antoniou:2017acq}, it has been appreciated that for a wide class of non-minimal couplings, the model accommodates \textit{both} scalarised BHs and the standard vacuum General Relativity (GR) solutions 
(see also \cite{Stefanov:2007eq,Doneva:2010ke,Antoniou:2017hxj}).
 This led to the conjecture that a phenomenon of ``spontaneous scalarisation" occurs in these models~\cite{Doneva:2017bvd,Silva:2017uqg}, akin to the spontaneous scalarisation of neutron stars first discussed in~\cite{Damour:1993hw}, within scalar tensor theories, but with the key difference that the phenomenon is triggered by strong gravity rather than by matter. Indeed, for some choices of the function defining the non-minimal coupling, scalarised BHs are thermodynamically preferred over the GR solutions and linearly stable~\cite{Blazquez-Salcedo:2018jnn}, suggesting such spontaneous scalarisation occurs dynamically.  

%\bigskip

Black hole spontaneous scalarisation was indeed confirmed to occur dynamically in a different class of model~\cite{Herdeiro:2018wub}, albeit within the same universality class related to non-minimal couplings. In this case, the scalar field non-minimally couples to the Maxwell invariant, rather than to higher curvature invariants, leading to a phenomenon of spontaneous scalarisation for electro-vacuum GR BHs. The technical simplification resulting from the absence of the higher curvature terms, allowed, besides showing dynamically the occurrence of spontaneous scalarisation, to explore the space of non-spherical static solutions that the model contains.\footnote{Charged scalarised BHs were also recently considered in models with a scalar field non-minimally coupled to higher curvature corrections~\cite{Doneva:2018rou}.}

%\bigskip

In all cases mentioned above, only static scalarised solutions have been considered, which connect to the Schwarzschild (or Reissner-Nordstr\"om) solution in the vanishing scalar field limit. It would be of interest to include rotation in this analysis, by considering the scalarisation of the Kerr solution. This is, however, considerably more challenging than the spherically symmetric cases, due to the combination of the higher curvature corrections and the higher co-dimensionality of the problem, due to the smaller isometry group. 

%\bigskip

In this paper we analyse the scalarisation of a different generalisation of the Schwarzschild solution, the Taub-Newman-Tamburino-Unti  solution (Taub-NUT)~\cite{Taub:1950ez,NUT}. This solution of the Einstein vacuum field equations can be regarded as the Schwarzschild solution with a NUT ``charge"  $n$ (hereafter referred to as the Schwarzschild-NUT solution) which endows the spacetime with rotation in the sense of promoting dragging of inertial frames. It is often described as a 'gravitational dyon' with both ordinary and magnetic mass.
The NUT charge $n$ plays a dual role to ordinary ADM mass $M$,  in much the same way that electric 
and magnetic charges are dual within Maxwell's electromagnetism~\cite{dam}. The Schwarzschild-NUT solution is not asymptotically flat in the usual sense, although it does obey the required fall-off conditions~\cite{misner-book}; for scalarisation, nonetheless, which is mostly supported in the neighbourhood of the horizon, the non-standard asymptotics are of lesser importance. In fact, the existence of scalarised solutions with these asymptotics supports the universality of the scalarisation phenomenon. The Schwarzschild-NUT solution is a case study in GR,  with a number of other unusual properties~\cite{misner-book}, and its Euclideanised version has been suggested to play a role in the context of quantum gravity~\cite{Hawking:ig}.

%\bigskip

Here, we shall exhibit  some basic properties of scalarised Schwarzschild-NUT solution. This scalarisation will have a geometric origin, occurring due to a non-minimal coupling of a scalar field and a higher curvature invariant. We shall consider two cases. The first case is the Einstein-Gauss-Bonnet-scalar (EGBs) model,
which uses the four dimensional Euler density (Gauss-Bonnet invariant). The corresponding solutions described herein generalise for a non-zero NUT parameter the recently discovered scalarised Schwarzschild BHs in~\cite{Doneva:2017bvd,Silva:2017uqg,Antoniou:2017acq}. The Kerr-like form of the Schwarzschild-NUT spacetime also allows, moreover, considering a second case, the Einstein-Chern-Simons-scalar (ECSs) model, wherein  the geometric scalarisation occurs due to a coupling between the scalar field and the Potryagin density.

%\bigskip

This paper is structured as follows. In Section~\ref{section2} we review the basic framework, including the field equations and ansatz. 
The main results of this work are given in Section~\ref{section3}. First, we present a test field analysis, wherein the backreaction of the scalar field on the metric is neglected. The corresponding solutions -  \textit{scalar clouds} - occur at branching points of the Schwarzschild-NUT solution, effectively requiring a quantisation of the parameters of this solution. Then the non-perturbative solutions are also studied. In particular, we exhibit their domain of existence in terms of the relevant parameters.  In Section~\ref{section4} we summarise our results and provide some further remarks together with possible avenues for future research.

%%%%%%%%%%%%%%%%%%%%%%%%%%%%%%%%%%%%%%%%%%%%%%%%%%%%%%%%%%%%%%%%%%%%%%%%%%%%%%%
\section{The model}
\label{section2}
%%%%%%%%%%%%%%%%%%%%%%%%%%%%%%%%%%%%%%%%%%%%%%%%%%%%%%%%%%%%%%%%%%%%%%%%%%%%%%%  

%%%%%%%%%%%%%%%%%%%%%%%%%%%%%%%%%%%%%%%%%%%%%%%%%%%%%%%%%%%%%%%%%%%%%%%%%%%%%%%
\subsection{The action and field equations}
%%%%%%%%%%%%%%%%%%%%%%%%%%%%%%%%%%%%%%%%%%%%%%%%%%%%%%%%%%%%%%%%%%%%%%%%%%%%%%% 

We consider the following action functional
\begin{eqnarray} 
\label{generalaction}
\mathcal{S}=
\frac{1}{16\pi}\int d^4x\sqrt{-g}
\left[
R-\frac{1}{2}\partial_\mu \phi\partial^\mu \phi +f(\phi) 
{\cal I}(g)
%+L_{\psi}
\right] \ ,
\end{eqnarray}
that describes a generalised gravitational theory containing
the Ricci scalar curvature $R$ and a massless real scalar field
$\phi$
which is non-minimally coupled to the spacetime geometry, via a coupling function $f(\phi) $ and a source term  ${\cal I}(g)$, which is built in terms of the metric tensor and  its derivatives only.

We shall consider two cases for  ${\cal I}( g)$,
corresponding to two different models:\footnote{
  ${\,^\ast\!}R^a{}_b{}^{cd}$ is the Hodge dual of the Riemann-tensor
\be
\label{Rdual}
{^\ast}R^a{}_b{}^{cd}\equiv \frac12 \eta^{cdef}R^a{}_{bef}\,,
\ee
 where  $\eta^{cdef}$ is 	the 4-dimensional Levi-Civita tensor,
	  $\eta^{cdef}=\epsilon^{cdef}/\sqrt{-g}$ and $\epsilon^{cdef}$ the Levi-Civita tensor density.
}
\begin{description} 
\item[{\bf i)}] the Einstein-Gauss-Bonnet-scalar (EGBs) model: ~ ${\cal I}={\mathcal L}_{GB}\equiv R^2 - 4 R_{ab}R^{ab} 
            + R_{abcd}R^{abcd}$ \ ,
\item[{\bf ii)}] the Einstein-Chern-Simons-scalar (ECSs) model:~~ ${\cal I}=\mathcal{L}_{CS}\equiv R \tilde R = {\,^\ast\!}R^a{}_b{}^{cd} R^b{}_{acd}$ \ .
\end{description}
Both 
${\mathcal L}_{GB}$
and
${\mathcal L}_{CS}$ are topological (total derivatives) in four dimensions.
Due to the coupling to the scalar field $\phi$ specified by $f(\phi)$, however, they become dynamical,
contributing to the equations of motion.

The field equations obtained by varying~(\ref{generalaction}) with respect
to  $\phi$ and $g_{ab}$ are
\be 
\label{KGeq}
\nabla^2 \phi+  \frac{df(\phi)}{d\phi}{\cal I}=0\ ,
\ee
for the scalar field, and
\begin{eqnarray} 
\label{grav-eqs}
 G_{ab}=\frac{1}{2}T_{ab}^{\rm (eff)} \ ,~ 
\end{eqnarray}
for the  gravitational field.
We have chosen to present (\ref{grav-eqs})  in a GR-like form, where  $G_{ab}$ is the Einstein tensor and $T_{ab} ^{\rm (eff)}$ is the \textit{effective} energy momentum tensor,
\begin{eqnarray}
\label{Teff}  
T_{ab} ^{\rm (eff)}\equiv  T_{ab}^{(\phi)}+  T_{ab}^{\rm (grav)}\  ,
\end{eqnarray}
which is a combination of the scalar field energy-momentum tensor, 
\be
\label{Tab-theta}
T_{ab}^{(\phi)}
\equiv    \left(\nabla_{a} \phi \right) \left(\nabla_{b} \phi \right) 
    -  \frac{1}{2}  g_{a b}  \left(\nabla_{c} \phi \right) \left(\nabla^{c} \phi\right)    \ ,
\ee
and a supplementary contribution which depends on the explicit form of ${\cal I}$.
For the cases herein we have
\begin{eqnarray}
\label{Teff-GB} 
 T_{ab}^{\rm (grav)}=
-\left(
g_{ak} g_{bj}+g_{bk} g_{aj}
\right)
\eta^{khcd}
\eta^{ijef}
R_{cdef}\nabla_h \nabla_i f(\phi) \ ,
\end{eqnarray}
%(with
%$
%\eta^{abcd}=\epsilon^{hbcd}/\sqrt{-g})
%$
 for a GB source term,
and  
\begin{eqnarray}
\label{Teff-CS} 
 T_{ab}^{\rm (grav)}=-8 C_{ab}\ , \qquad {\rm with}~~
C^{ab} = [\nabla_c f(\phi)]
\epsilon^{cde(a}\nabla_e R^{b)}{}_d+
[\nabla_{c}\nabla_{d} f(\phi)]
{\,^\ast\!}R^{d(ab)c}\ ,
\end{eqnarray}
in the CS case.

%%%%%%%%%%%%%%%%%%%%%%%%%%%%%%%%%%%%%%%%%%%%%%%%%%%%%%%%%%%%%%%%%%%%%%%%%%%%%%%
\subsection{The spontaneous scalarization mechanism}
%%%%%%%%%%%%%%%%%%%%%%%%%%%%%%%%%%%%%%%%%%%%%%%%%%%%%%%%%%%%%%%%%%%%%%%%%%%%%%% 

The possible occurrence of a  `spontaneous scalarisation' in this class of models relies on two ingredients. Firstly, there must exist a \textit{scalar-free solution}, with
\begin{eqnarray}
 \phi=\phi_0 \ ,
\end{eqnarray}
which is the fundamental solution of  equation (\ref{KGeq}).
This requirement implies the coupling function should satisfy the condition
\begin{eqnarray}
\label{cond}
\frac{\partial f}{\partial \phi}\Big |_{\phi=\phi_0}=0 \ .
\end{eqnarray}
This condition signifies that the usual GR solutions also solve  the considered model, forming the  fundamental set. We remark one can set $\phi_0=0$ without any loss of generality (via a field redefinition). 

Secondly, the model possesses a second set of solutions, with a nontrivial scalar field -- {\it the scalarised solutions}.
In the asymptotically flat case, these solutions can be entropically preferred
over the fundamental ones
 ($i.e.$ they maximise the entropy for given global charges)~\cite{Blazquez-Salcedo:2018jnn,Herdeiro:2018wub}.
Moreover, they are smoothly connected with the scalar-free solution, approaching it for $\phi=0$.

%%%%%%%%%%%%%%%%%%%%%%%%%%%%%%%%%%%%%%%%%%%%%%%%%%%%%%%%%%%%%%%%%%%%%%%%%%%%%%%
\subsection{The scalar-free Schwarzschild-NUT solution}
%%%%%%%%%%%%%%%%%%%%%%%%%%%%%%%%%%%%%%%%%%%%%%%%%%%%%%%%%%%%%%%%%%%%%%%%%%%%%%% 

The scalar-free solution we wish to consider, within the model~(\ref{KGeq})-(\ref{grav-eqs}), is the Schwarzschild-NUT solution. It has $\phi=0$ and line element:
\begin{eqnarray}
\label{metric}
ds^2=-N(r)\sigma^2(r)(dt + 2n\cos \theta  d\varphi)^2+\frac{dr^2}{N(r)}+g(r)(d\theta^2+\sin^2 \theta d\varphi^2)~,
\end{eqnarray}  
with 
\begin{eqnarray}
\label{TN-g}
g(r)=r^2+n^2  \ ,
\end{eqnarray}
and
\begin{eqnarray}
\label{TN}
N(r)=1-\frac{2(M r+n^2)}{r^2+n^2}\ , \qquad \sigma(r)=1 \ .
\end{eqnarray}
In (\ref{metric}), $\theta$ and $\varphi$ are the standard angles parameterising 
an $S^2$ with the usual range. 
As usual,
we define the NUT parameter
 $n$
(with $n\geqslant 0$, without any loss of generality),
in terms of the coefficient appearing in the differential
$dt+2n\cos \theta  d\varphi$.
Also, $r$ and $t$ are the `radial' and `time' coordinates.

This metric  possesses a  horizon located at\footnote{For $n\neq 0$,
 a negative value  of the 'electric' mass $M$ is 
allowed for the NUT solution. 
Such configurations are found for $0<r_h<n$ and do not possess a Schwarzschild limit.
As such, they are ignored in this work.}
\begin{eqnarray}
\label{rH}
 r_h=M+\sqrt{M^2+n^2}>0.
\end{eqnarray} 
As in the Schwarzschild limit, $N(r_h) = 0$ is only a coordinate singularity where all curvature invariants are finite. 
In fact, a nonsingular extension across this null surface can be found~\cite{Hawking:1973uf}.

Both the Gauss-Bonnet and Pontryagin densities are non-vanishing in this background:
\begin{equation} 
\label{LGB0}
{\mathcal L}_{GB}=
\frac{48 M^2}{g^6}[g^2-16 n^2 r^2](r^2-n^2)
\left[
1+\frac{n}{M}\frac{(r-n)(g+4nr)}{(r+n)(g-4nr)}
\right]
\left[
1-\frac{n}{M}\frac{(r+n)(g-4nr)}{(r-n)(g+4nr)}
\right] \ ,
\end{equation}
and
\begin{eqnarray} 
\label{LCS0}
\mathcal{L}_{CS}=
\frac{24 n}{r g^2}
\left[1-\frac{N(r^2-3n^2)}{g} \right]
\left[1-\frac{N(3r^2-n^2)}{g} \right] \ .
\end{eqnarray}
In particular, the vanishing of  $\mathcal{L}_{CS}$ for $n=0$, justifies the interpretation of $n$ as a gravitomagnetic ``mass".
 
We remark that the  thermodynamical description of (Lorentzian signature)  solutions with NUT charge  
is still poorly understood. Consequently, the solutions' properties in this respect will not be pursued herein 
($e.g.$ the usual entropy comparison between the vacuum and scalarised solutions).
Still, 
the  mass of solutions with NUT charge can be computed by employing 
the quasi-local formalism supplemented with the boundary counter-term method
\cite{Astefanesei:2006zd}.
A direct computation shows that, similar to the Einstein gravity case,
the mass of the solutions
is identified with the constant $M$ in the far field expansion of
the metric function  $g_{tt}$,
\begin{eqnarray}
\label{Mass}
g_{tt}=-N\sigma^2=-1+\frac{2M}{r}+\dots~.
\end{eqnarray} 

%%%%%%%%%%%%%%%%%%%%%%%%%%%%%%%%%%%%%%%%%%%%%%%%%%%%%%%%%%%%%%%%%%%%%%%%%%%%%%%
\subsection{Ansatz and choice of coupling}
%%%%%%%%%%%%%%%%%%%%%%%%%%%%%%%%%%%%%%%%%%%%%%%%%%%%%%%%%%%%%%%%%%%%%%%%%%%%%%% 

To consider the scalarised  generalisations of the Schwarzschild-NUT spacetime, we consider again the metric ansatz~(\ref{metric}).
The metric functions $N(r),~\sigma(r)$ and $g(r)$ are now unknowns, that shall be determined from the requirement that this  metric solves the field equations~(\ref{KGeq})-(\ref{grav-eqs}). We have kept some metric gauge freedom in (\ref{metric}),  which shall be conveniently fixed later.

The scalar field is a function of $r$ only,
\begin{eqnarray}
\label{scalar}
 \phi \equiv \phi(r).
\end{eqnarray} 
In this work, we shall restrict 
ourselves
to the simplest coupling
function which satisfies  the condition (\ref{cond}) (with $\phi_0=0$):
\begin{eqnarray} 
\label{f0}
f(\phi)=\alpha \phi^2 \ ,
\end{eqnarray}
with the coupling constant $\alpha$ an input parameter.

The equations satisfied by  $N(r),\sigma(r),g(r)$ and $\phi(r)$ 
are rather involved and unenlightening; we shall not include them here. We notice, however, that they can also be derived from the following effective action\footnote{In these expressions
 a prime denotes a derivative $w.r.t.$ the radial coordinate $r$.}:
\begin{eqnarray}
\label{Leff}
{\cal L}_{\rm eff}={\cal L}_E+{\cal L}_\phi
+
\alpha
{\cal L}_{ {\cal I}\phi}~,
\end{eqnarray} 
where
\begin{eqnarray}
\label{LE}
\nonumber
&&
 {\cal L}_E=
2\sigma
\left[
1+\left(\frac{N'}{2N}+\frac{g'}{4g}+\frac{\sigma'}{\sigma} \right)Ng'+\frac{\sigma^2 N}{g}n^2
\right] \ ,
\qquad
{\cal L}_{\phi}=-\frac{1}{2}N\sigma g \phi'^2~,
\end{eqnarray}
and ${\cal L}_{ {\cal I}\phi}={\cal L}_{ GB \phi}$  or ${\cal L}_{ {\cal I}\phi}={\cal L}_{ CS \phi}$,
where
\begin{eqnarray}
\label{LGBs}
&&
\nonumber 
{\cal L}_{GB \phi }=
8  \sigma   N 
\left[
\left( 
\frac{N'}{N}+\frac{2\sigma'}{\sigma}
\right)
\left(
1-\frac{Ng'^2}{4g}
\right)
+\frac{n^2 N\sigma^2}{g}
\left(
\frac{3N'}{N}-\frac{2g'}{g}+\frac{6\sigma'}{\sigma}
\right)
				\right]
				\phi \phi' \ ,
\end{eqnarray}
and
\begin{eqnarray}
\label{LCSs}
&&
\nonumber 
{\cal L}_{CS \phi }=
16 n N^2 \sigma^2
\left[
\frac{1}{4}
\left(
\frac{N'}{N}-\frac{g'}{g}+\frac{4\sigma'}{\sigma}
 \right)
\left(
\frac{N'}{N}-\frac{g'}{g}
\right)
+\frac{\sigma'^2}{\sigma^2}
-\frac{1}{g N}
-\frac{2n^2\sigma^2}{g^2}
				\right]
				\phi \phi' \ .
\end{eqnarray}
Remarkably, one can see that,
due to the factorisation of the angular dependence permitted by the
 metric ansatz (\ref{metric}),
 all functions solve
$second$ order equations of motion \textit{also} in ECSs case\footnote{Without this factorisation, the metric functions
would
solve third order partial differential equations, as in the case of the Kerr metric in ECSs theory,
with $f(\phi)=\alpha \phi$ \cite{Delsate:2018ome}. }.

The reduced action (\ref{Leff}) makes transparent the residual scaling symmetry of the problem 
($\lambda$ is a nonzero constant)
\begin{eqnarray}
\label{scaling}
\alpha \to \alpha \lambda^2,~~
r\to \lambda r,~~
n\to \lambda n,~~
{\rm and}~~~
g\to \lambda^2 g,
\end{eqnarray}
%which can be used to fix the value of $\alpha$ or $n$.
%As such the model has only one free input parameter.
 corresponding to the freedom
in choosing a unit length. 
Therefore the solutions are  characterised
by dimensionless quantities such $\alpha/M^2$, $Q_s/M$ and $n/M$, which shall be used below to describe them
($Q_s, M$ shall be defined below, $cf.$~Eq. (\ref{inf}).

%%%%%%%%%%%%%%%%%%%%%%%%%%%%%%%%%%%%%%%%%%%%%%%%%%%%%%%%%%%%%%%%%%%%%%%%%%%%%%%
\section{Scalarised solutions}
\label{section3}
%%%%%%%%%%%%%%%%%%%%%%%%%%%%%%%%%%%%%%%%%%%%%%%%%%%%%%%%%%%%%%%%%%%%%%%%%%%%%%% 

%%%%%%%%%%%%%%%%%%%%%%%%%%%%%%%%%%%%%%%%%%%%%%%%%%%%%%%%%%%%%%%%%%%%%%%%%%%%%%%
\subsection{The zero-mode }
%%%%%%%%%%%%%%%%%%%%%%%%%%%%%%%%%%%%%%%%%%%%%%%%%%%%%%%%%%%%%%%%%%%%%%%%%%%%%%% 
 We
first consider the test field limit.  That is,  we focus on the scalar field equation (\ref{KGeq}) on a fixed Schwarzschild-NUT background: 
\begin{eqnarray} 
\label{KG0}
 \frac{1}{g(r)}\frac{d}{dr}
\left[
g(r)N(r) \phi'
\right]
+2\alpha \phi {\cal I}=0 \ ,
\end{eqnarray}
where
${\cal I}$
is
given by (\ref{LGB0}) or (\ref{LCS0}), respectively. For our purpose, solving (\ref{KG0}) is an eigenvalue problem: selecting appropriate boundary conditions and fixing the values of $(n,\alpha)$,  selects
a discrete set of BHs, as specified by the mass parameter $M$.
These are the bifurcation points
of the scalar-free solutions. The (test) scalar field profiles
they support - hereafter scalar clouds - are distinguished
by the number of nodes of $\phi$, $k\geqslant 0$.

%%%%%%%%%%%%%%%%%%%%%%%%%%%%%%%%%%%%%%%%%%%%%%%%%%%%%%%%%
\begin{figure}[ht!]
\begin{center}
%\subfigure[Mass]
{\label{c2}\includegraphics[width=8cm]{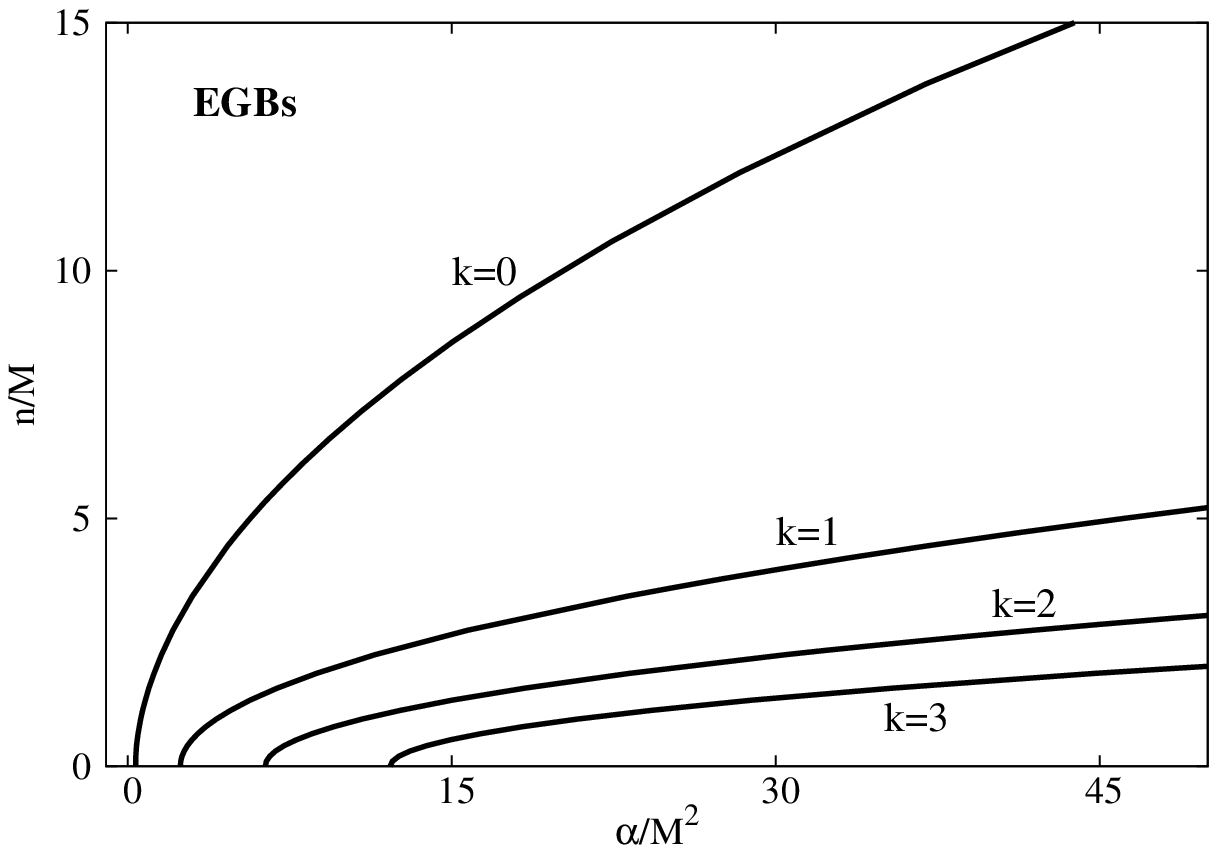}}
%\subfigure[$\omega, M/Q$]
{\label{ss0}\includegraphics[width=7.8cm]{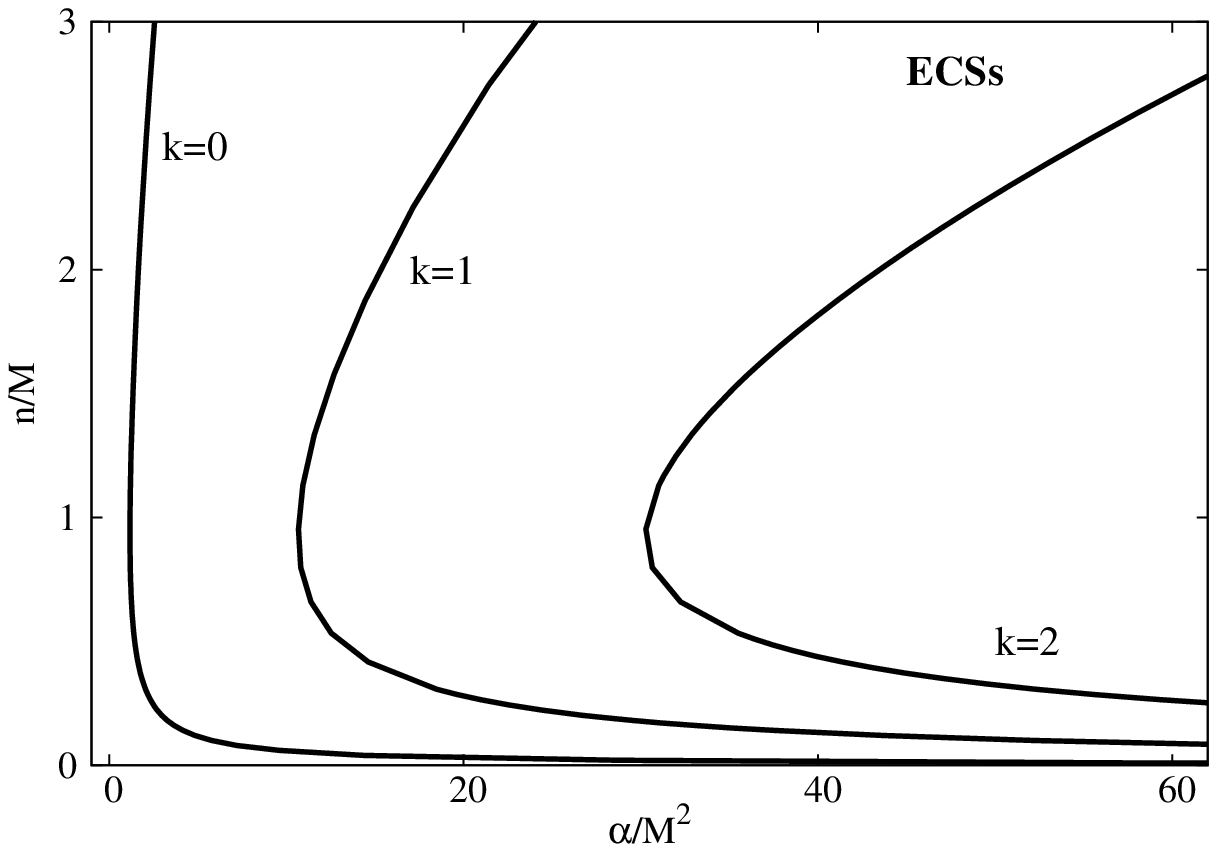}}
%%%\subfigure[$\omega, M/Q$]
%%\end{center}
\caption{The $k=0,1,2$ existence lines for the  EGBs (left panel) and the ECSs (right panel) models in a NUT charge $vs.$ coupling parameter diagram.   
\label{existence-lines}
}
\end{center}
\end{figure} 
%%%%%%%%%%%%%%%%%%%%%%%%%%%%%%%%%%%%%%%%%%%%%%%%%%%%%%%%%

%%%%%%%%%%%%%%%%%%%%%%%%%%%%%%%%%%%%%%%%%%%%%%%%%%%%%%%%%
\begin{figure}[ht!]
\begin{center}
%\subfigure[Mass]
{\label{c2new}\includegraphics[width=8cm]{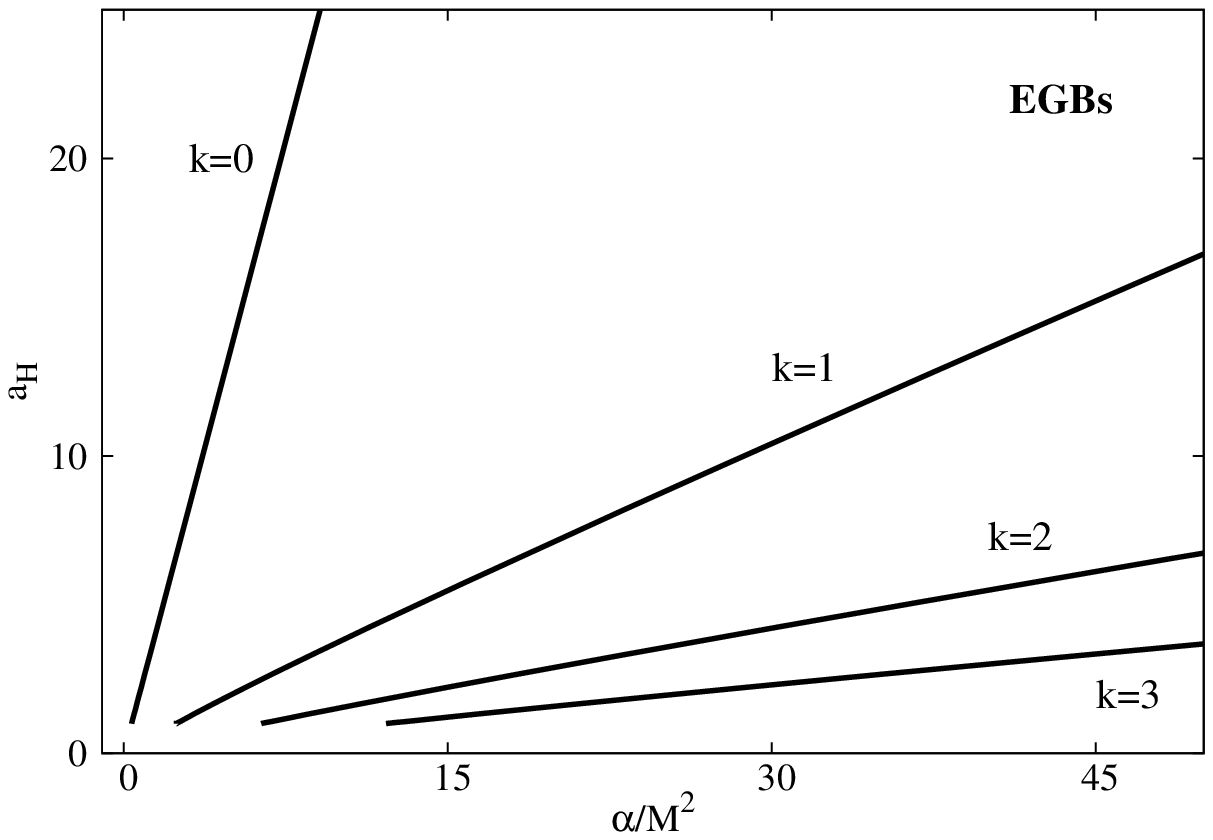}}
%\subfigure[$\omega, M/Q$]
{\label{ss0new}\includegraphics[width=7.8cm]{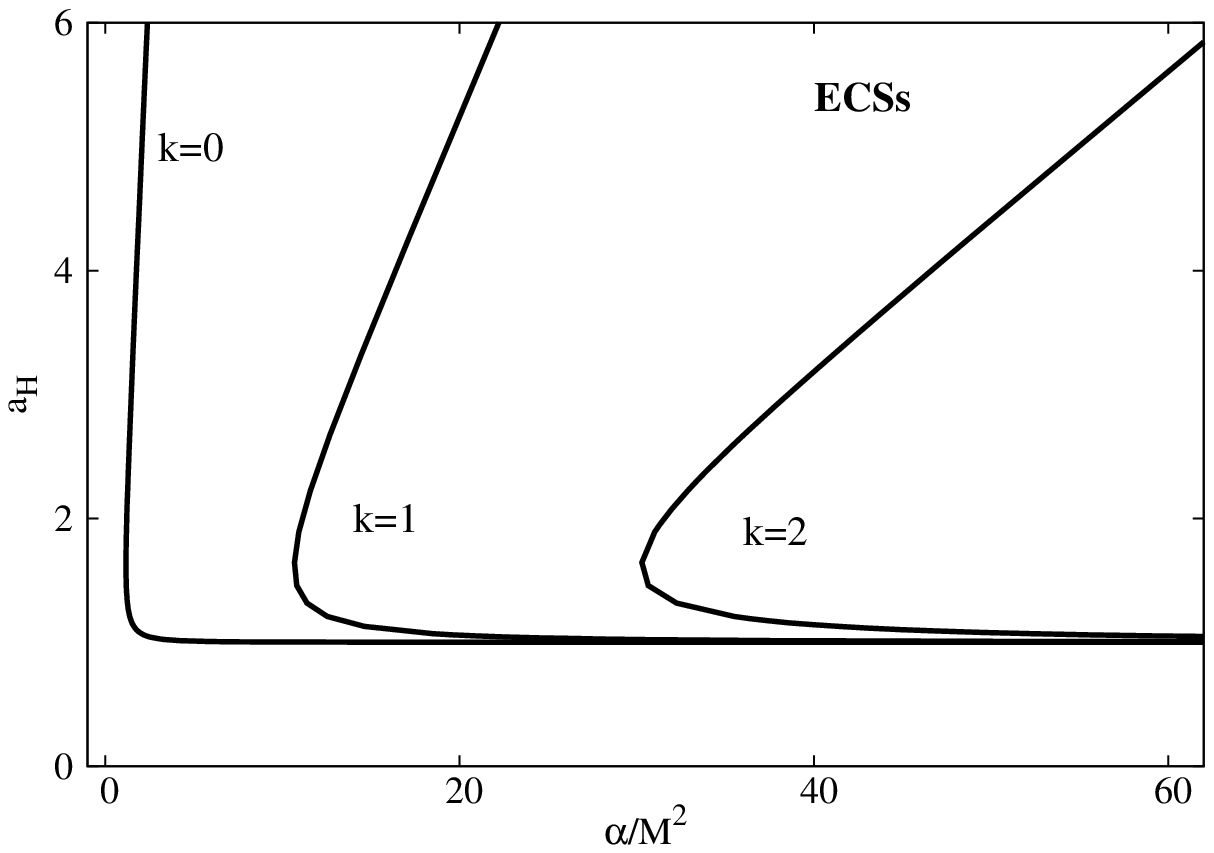}}
%%%\subfigure[$\omega, M/Q$]
%%\end{center}
\caption{The $k=0,1,2$ existence lines for the  EGBs (left panel) and the ECSs (right panel) 
models in a reduced horizon area $a_H=A_H/(16\pi M^2)$ $vs.$ coupling parameter diagram, where $A_H\equiv 4\pi g(r_h)$ is the area of the round-$S^2$ part of the line element (\ref{metric})
evaluated at the horizon.   
\label{existence-lines-aH}
}
\end{center}
\end{figure} 
%%%%%%%%%%%%%%%%%%%%%%%%%%%%%%%%%%%%%%%%%%%%%%%%%%%%%%%%%

The boundary behaviours of the scalar field are regularity 
on the horizon and vanishing at infinity. More concretely, 
the approximate expression of the solution close to the horizon reads 
\begin{eqnarray}
\label{ss1}
\phi(r) =\phi_0+\phi_1(r-r_h)+\mathcal{O}(r-r_h)^2 \ , 
\end{eqnarray}
with
\begin{eqnarray}
\label{ss2}
 \phi_1=\frac{24\alpha \phi_0(n^2-r_h^2)}{r_h(r_h^2+n^2)^2} \qquad {\rm for~EGBs} \qquad {\rm and} \qquad 
\phi_1=-\frac{48\alpha \phi_0 n}{(r_h^2+n^2)^2} \qquad {\rm for~ECSs} \ .
\end{eqnarray}
At infinity one finds
\begin{eqnarray}
\label{sinf}
\phi(r) = \frac{Q_s}{r}+\frac{Q_s(r_h^2-n^2)}{2r_h r^2}+ \dots,
\end{eqnarray}
in both cases,
with $Q_s$ a constant fixed by numerics. No exact solution appears to  exist even for $n=0$ and  equation (\ref{KG0}) is solved numerically. The corresponding results - the \textit{existence lines} - for the EGBs and ECSs are shown in Figure \ref{existence-lines},
in units given by the mass parameter  $M$. Amongst the notable features we highlight: $(i)$ the existence in both cases,  of a minimal value of the coupling  constant $\alpha$
below which no zero modes with a given number of nodes exists\footnote{This holds also the $n=0$ limit of the GB model;
there one finds $\alpha/M^2=0.3631$.}; $(ii)$ the CS model possess a double branch structure,
with the existence of two different critical
NUT solutions for the same $(\alpha,M)$,
which are distinguished by the value of the NUT charge.

In Figure \ref{existence-lines-aH}, the same existence lines are shown in terms of the reduced horizon area. One observes that the larger the horizon grows, the stronger the coupling has to be at the branching point, for the EGBs case. For the ECSs case, on the other hand, both this and the opposite trend can be found, corresponding to the two branches.

%%%%%%%%%%%%%%%%%%%%%%%%%%%%%%%%%%%%%%%%%%%%%%%%%%%%%%%%%%%%%%%%%%%%%%%%%%%%%%%
\subsection{Non-perturbative results}
%%%%%%%%%%%%%%%%%%%%%%%%%%%%%%%%%%%%%%%%%%%%%%%%%%%%%%%%%%%%%%%%%%%%%%%%%%%%%%%

%%%%%%%%%%%%%%%%%%%%%%%%%%%%%%%%%%%%%%%%%%%%%%%%%%%%%%%%%%%%%%%%%%%%%%%%%%%%%%%
\subsubsection{Asymptotics and remarks on numerics}
%%%%%%%%%%%%%%%%%%%%%%%%%%%%%%%%%%%%%%%%%%%%%%%%%%%%%%%%%%%%%%%%%%%%%%%%%%%%%%%

We are interested in scalarised solutions whose far field asymptotics are similar,  to leading order,
to those the scalar-free solution (\ref{TN}):  $N(r)\to 1$, $g(r)\to r^2$, $\sigma(r)\to 1$
and $\phi(r)\to 0$, as $r\to \infty$.
The solution will also posses a horizon at $r=r_h>0$,
where
$N(r_h)=0$, and $g(r)$, $\sigma(r)$ are strictly positive.

Since the scalarised configurations are smoothly connected with the 
vacuum NUT solution, 
it is natural to 
choose the same metric gauge (\ref{TN-g}).\footnote{Most
of the ECSs solutions, however, were constructed for a  metric gauge choice
with $\sigma(r)=1$ and unknown metric functions $N,g$.
} 
Then we are left with a system of  three non-linear ODEs (plus a constraint)
for the functions $N,\sigma$ and $\phi$.
These equations do not appear to possess closed form solutions,
and are constructed by solving
numerically a boundary value problem
in the domain 
$r_h\leqslant r<\infty$ (with $r_h>0$). 
As $r\to r_h$, one takes a power series expansion of the solution in $(r-r_h)$,
the first terms  being
\begin{equation} 
\label{hor}
N(r)= N_1(r-r_h)+N_2(r-r_h)^2+\dots, \ \ \ 
\sigma(r)=\sigma_0+ \sigma_1(r-r_h)+\dots, \ \ \ 
~~
\phi(r)= \phi_0+ \phi_1(r-r_h)+\dots, 
\end{equation}
with all coefficients in this expansion being fixed by the set 
$\{N_1,\sigma_0,\phi_0 \}$.
These parameters are
subject to the following constraint:
\begin{eqnarray} 
\label{condGB}
&&
 N_1 r_h=1+\frac{96 \alpha^2 N_1^2(r_h^2-n^2 \sigma_0^2)\phi_0^2}{(n^2+r_h^2)^2} \ ,~~{\rm for~~EGBs} \ ,
\\
&&
\label{condCS}
N_1 r_h=1+\frac{192 \alpha^2 N_1^3 n^2 r_h \sigma_0^2 \phi_0^2}{(n^2+r_h^2)^2} \ ,~~{\rm for~~ECSs} \ .
\end{eqnarray}
The leading order expansion in the far field is the same in both cases:
\begin{eqnarray} 
\label{inf}
N(r)=1-\frac{2M}{r}-\frac{Q_s^2-8n^2}{4r^2}+\dots, \qquad
\sigma(r)=1-\frac{Q_s^2}{8r^2}+\dots, \qquad
\phi(r)=\frac{Q_s}{r}+\frac{MQ_s}{r^2}+ \dots,
\end{eqnarray}
 introducing the parameters $M$ and $Q_s$ which are fixed by numerics; $M$ is identified as the mass of solutions, $cf.$ (\ref{Mass}).
 The constant $Q_s$
can be interpreted as the 'scalar `charge', giving a quantitative measure of `hairiness' of the solutions.

For the EGBs model, the numerical integration was carried out using a standard
shooting method. In this approach, we evaluate the initial conditions at   
$r=r_h+ 10^{-6}$ for global tolerance $10^{-14}$, adjusting for shooting parameters
 and integrating
towards $r\to \infty$. 
The ECSs solutions were found by using a professional
software package \cite{COLSYS}. 
This solver employs a collocation method for boundary-value differential 
equations
and a damped Newton method of quasi-linearisation.  
In both cases,
the input parameters were $\{ r_h, n; \alpha \}$.
We also remark that
the configurations reported in this work are
regular on and outside the horizon,
all curvature invariants being finite.

%%%%%%%%%%%%%%%%%%%%%%%%%%%%%%%%%%%%%%%%%%%%%%%%%%%%%%%%%
\begin{figure}[ht!]
\begin{center}
%\subfigure[Mass]
{\includegraphics[width=8cm]{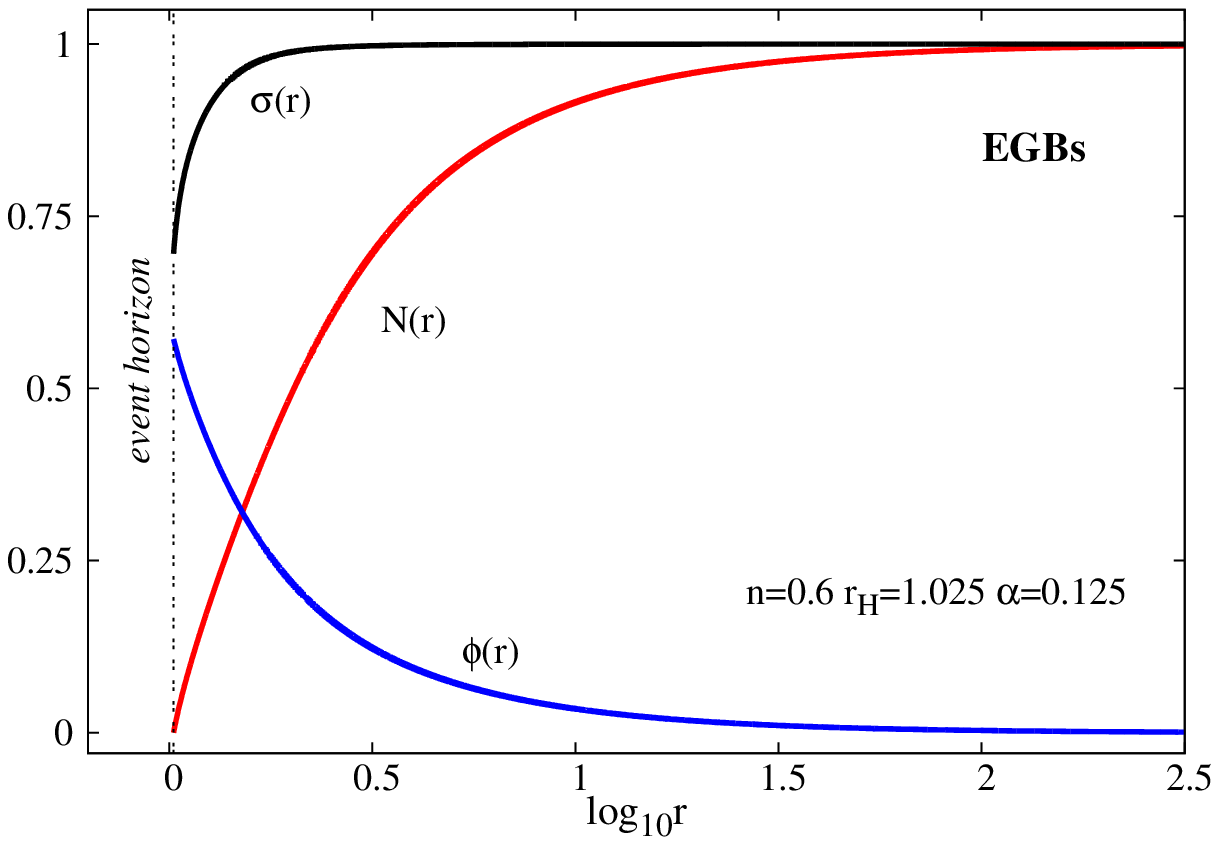}}
%\subfigure[$\omega, M/Q$]
{\includegraphics[width=7.8cm]{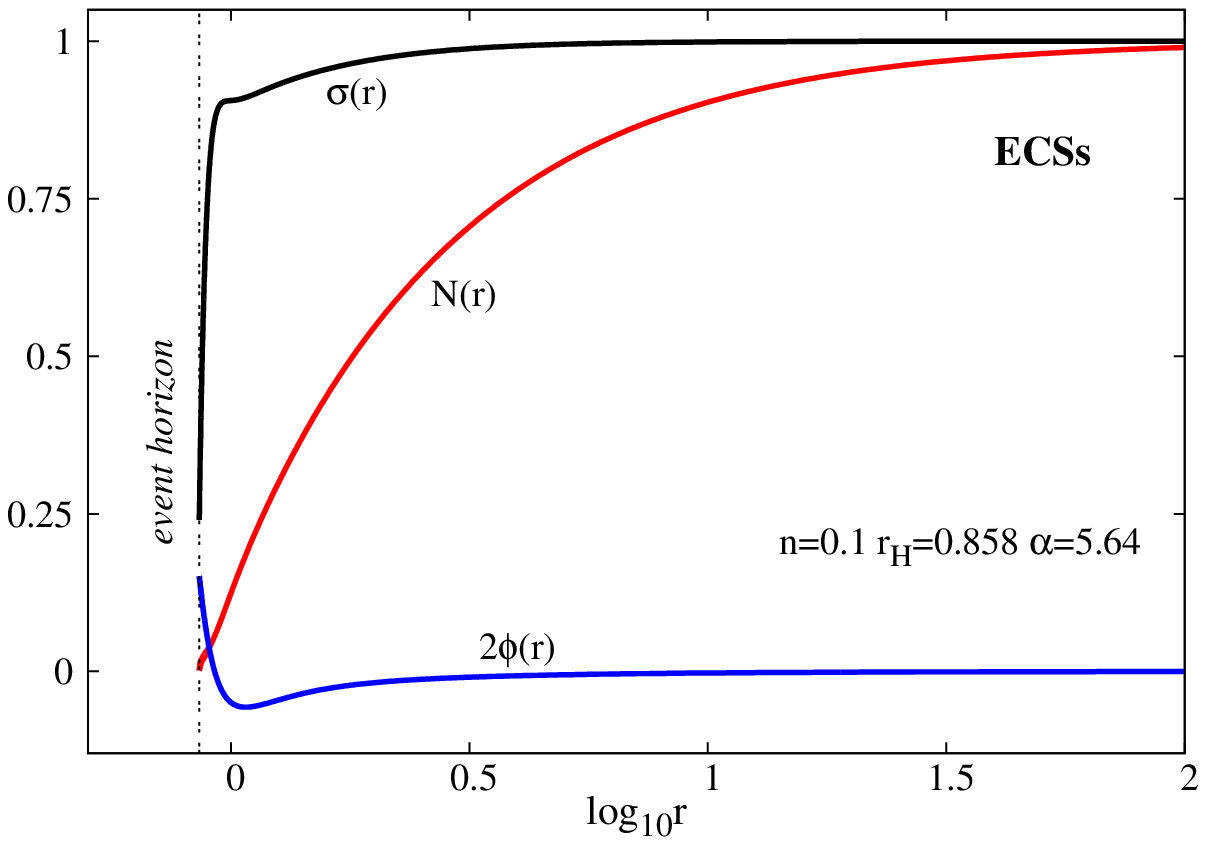}}
%%%\subfigure[$\omega, M/Q$]
%%\end{center}
\caption{Profile functions of a typical  EGBs scalarised nodeless solution (left panel) and of  a  typical  ECSs scalarised $k=1$ solution (right panel).}
\label{profile}
\end{center}
\end{figure} 
%%%%%%%%%%%%%%%%%%%%%%%%%%%%%%%%%%%%%%%%%%%%%%%%%%%%%%%%%

%%%%%%%%%%%%%%%%%%%%%%%%%%%%%%%%%%%%%%%%%%%%%%%%%%%%%%%%%%%%%%%%%%%%%%%%%%%%%%%
\subsubsection{Numerical results}
%%%%%%%%%%%%%%%%%%%%%%%%%%%%%%%%%%%%%%%%%%%%%%%%%%%%%%%%%%%%%%%%%%%%%%%%%%%%%%% 

As an illustration of the numerical results obtained in this way, the profile of a typical solution is shown in Figure \ref{profile}
for a nodeless scalar field in EGBs model and a $k=1$
 ECSs configuration.\footnote{For a given $k$, the typical profiles  are similar in both models.}

\begin{figure}[ht!]
\begin{center}
%\subfigure[Mass]
{\includegraphics[width=8cm]{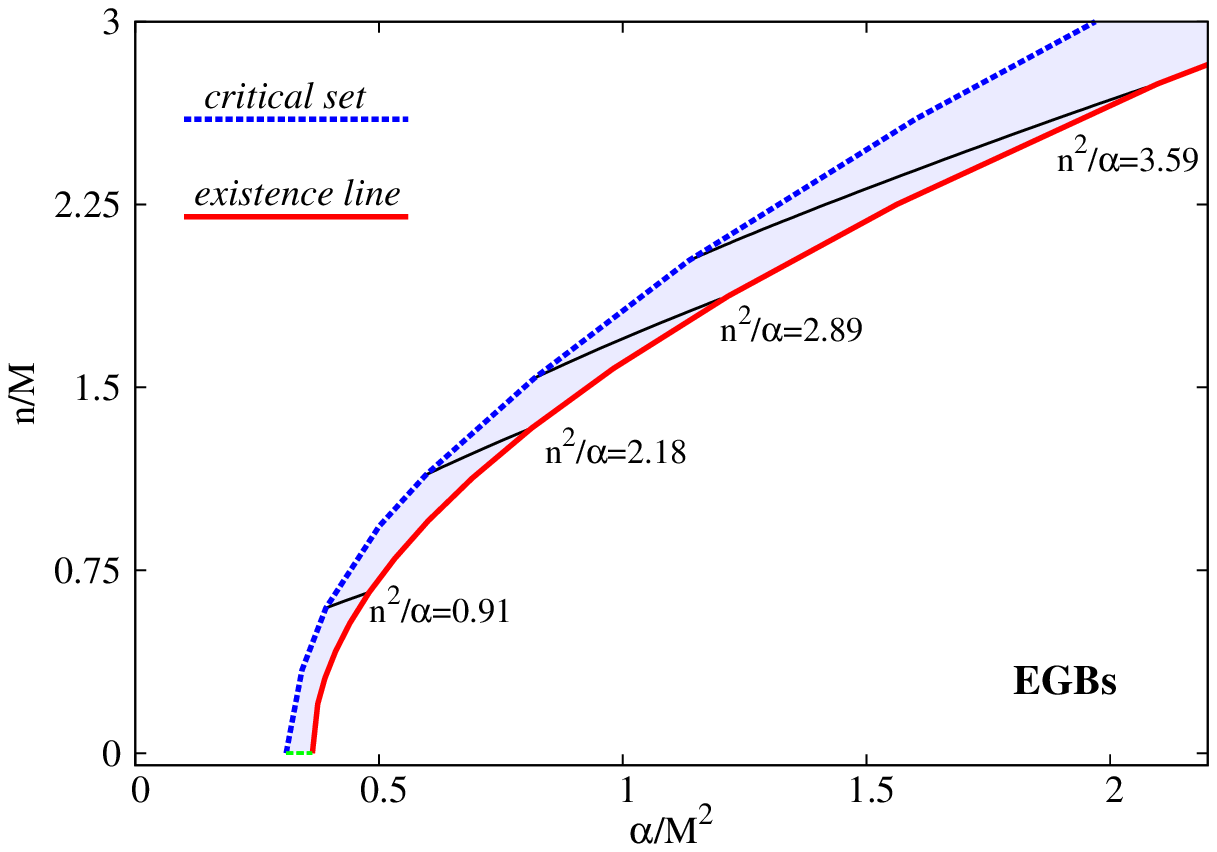}}
%\subfigure[$\omega, M/Q$]
{\includegraphics[width=7.8cm]{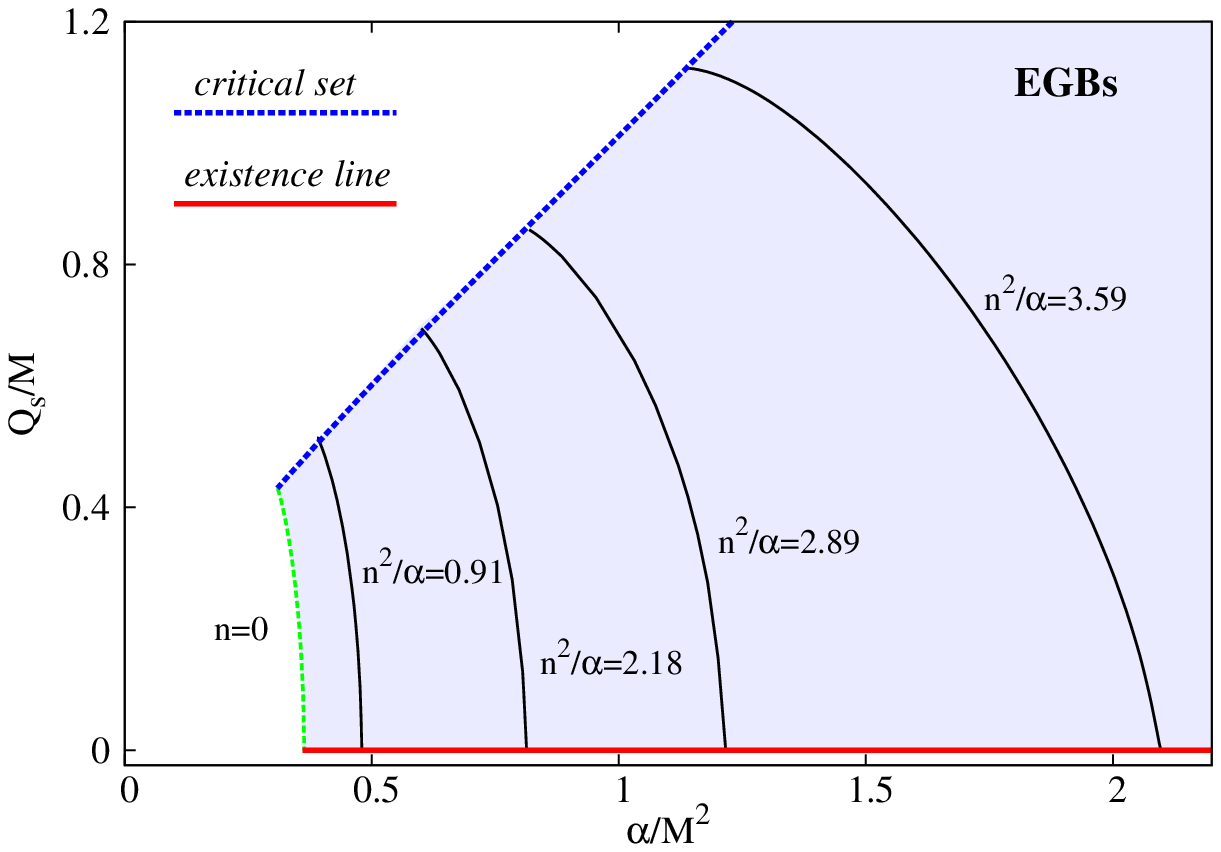}}
%%%\subfigure[$\omega, M/Q$]
%%\end{center}
\caption{
Part of the domain of existence in the $(\alpha,n)$ and $(\alpha,Q_s)$ spaces  
(in units set by $M$) 
where scalarized Schwarzschild-NUT solutions exist
in EGBs model.
Only the fundamental ($k=0$) scalarization region is exhibited, but  an infinite
number of domains, indexed by the node number of the scalar field, should exist.
}
\label{domainEGBs}
\end{center}
\end{figure}

Let us now take a closer look at the domain of existence of the scalarised Schwarzschild-NUT solutions, starting with the EGBs model. The corresponding domain of existence is shown in Figure \ref{domainEGBs}. The solutions form a band in the ($\alpha,n$)  (or $(\alpha,Q_s)$)  plane, bounded by two curves:
$(i)$ the existence line and $(ii)$ a set of critical solutions explained below. In units of $M$, for each $\alpha$ above a minimal value, scalarised solutions bifurcate from the  vacuum Schwarzschild-NUT solution with a particular value of $n$ at the existence line. Then,  for given $(n,M)$,
the scalarized solutions exist only within some $\alpha$ range,  $\alpha_{min}<\alpha<\alpha_{max}$,
with the limiting values increasing with the ratio $n/M$.
% for fixed $(\alpha, M)$,
%the scalarized solutions exist for small values of $n$ only,
Each constant $\alpha$-branch starts on the existence line and
ends at a critical  configuration where the numerics stops to converge.
The critical configuration does  not possess any special properties.
As with other solutions in various $d=4$ EGBs models,
its existence can be traced to the horizon condition (\ref{condGB}):
the reality of $N_1$ implies that 
solutions with given $(r_h,n)$ can only exist if the
coupling constant $\alpha$ is smaller than a critical value.

%%%%%%%%%%%%%%%%%%%%%%%%%%%%%%%%%%%%%%%%%%%%%%%%%%%%%%%%%
%%%%%%%%%%%%%%%%%%%%%%%%%%%%%%%%%%%%%%%%%%%%%%%%%%%%%%%%%
%
%%%%%%%%%%%%%%%%%%%%%%%%%%%%%%%%%%%%%%%%%%%%%%%%%%%%%%%%%
\begin{figure}[ht!]
\begin{center}
%\subfigure[Mass]
{\includegraphics[width=8cm]{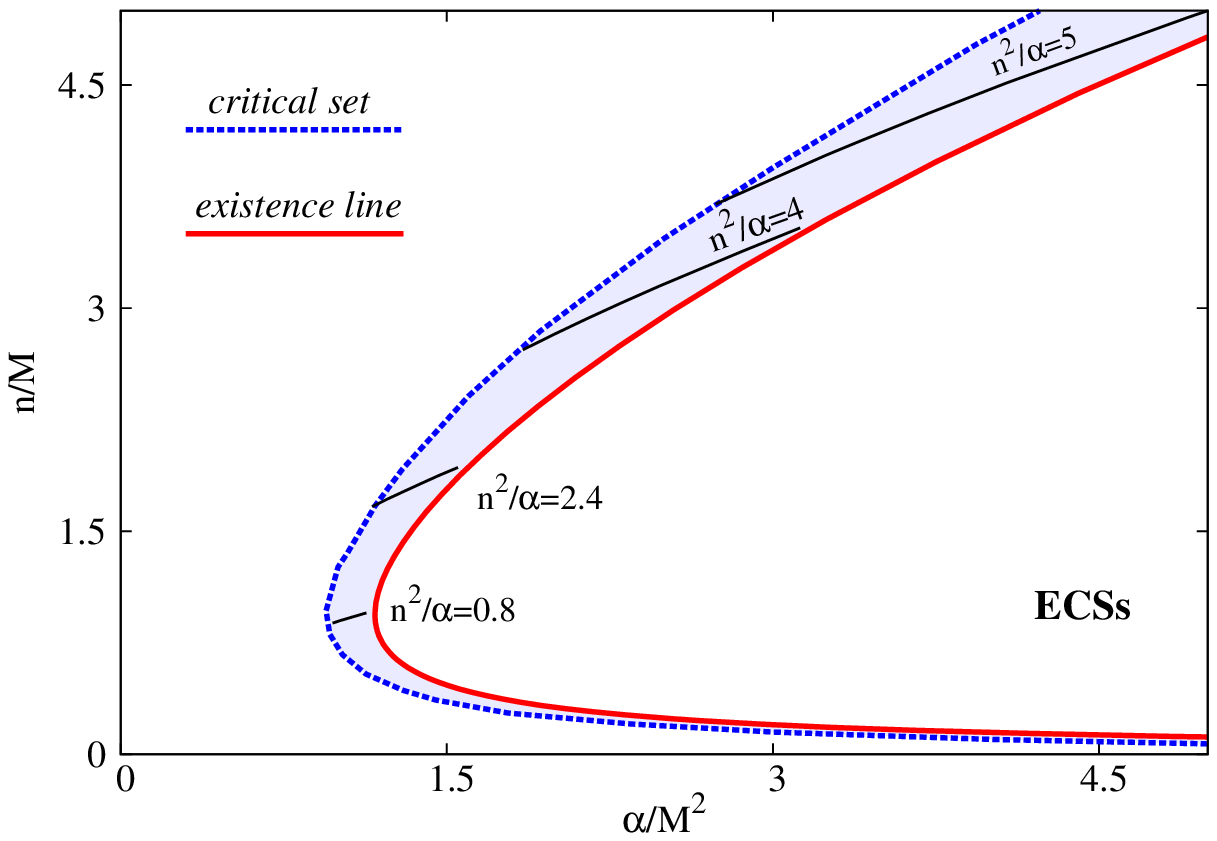}}
%\subfigure[$\omega, M/Q$]
{\includegraphics[width=7.8cm]{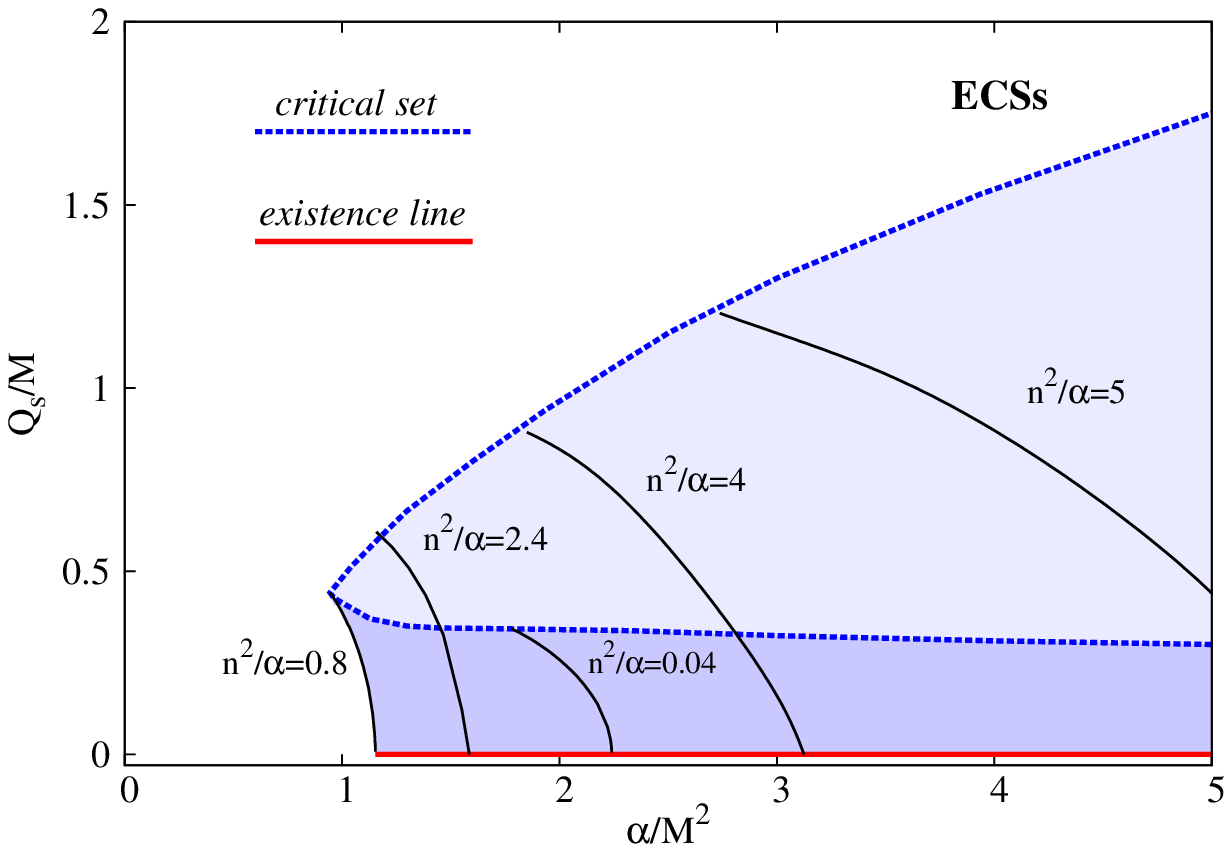}}
%%%\subfigure[$\omega, M/Q$]
%%\end{center}
\caption{Same as Figure  \ref{domainEGBs}
for scalarised Schwarzschild-NUT solutions of the ECSs model.}
\label{domainECSs}
\end{center}
\end{figure} 
%%%%%%%%%%%%%%%%%%%%%%%%%%%%%%%%%%%%%%%%%%%%%%%%%%%%%%%%%

A similar picture to what has just been described can be found for the ECSs case. 
As can be anticipated from Figure \ref{existence-lines}, however, 
there is a new ingredient: the existence of a double band structure of solutions -
see Figure  \ref{domainECSs}.
One can see that, for any $n$, no scalarized solutions exist below a critical value of $\alpha$.
Also, the $n\to 0$
limit cannot be reached for any finite $\alpha$, as expected from the fact that the Chern-Simons term vanishes for $n=0$. 
The domain of existence of the solutions is bounded again by  the existence line and  a critical line.
The presence of  critical ECSs configurations can be 
traced back, again, to the near horizon condition
(\ref{condCS}) which, for given  $(r_h,n)$, stopped being satisfied for large 
enough $\alpha$.

%%%%%%%%%%%%%%%%%%%%%%%%%%%%%%%%%%%%%%%%%%%%%%%%%%%%%%%%%%%%%%%%%%%%%%%%%%%%%%%
\section{Further remarks }
\label{section4}
%%%%%%%%%%%%%%%%%%%%%%%%%%%%%%%%%%%%%%%%%%%%%%%%%%%%%%%%%%%%%%%%%%%%%%%%%%%%%%% 
The main purpose of this work was to investigate the basic properties of the scalarised Schwarzschild-NUT solution, viewed as a 
toy model for the scalarised rotating Kerr BH. 
Indeed, the  line-element of the former solution is Kerr-like, in the sense that it has a crossed  metric component $g_{\varphi t}$, $cf.$ (\ref{metric}). This term does not produce an ergoregion but it leads to
an effect similar to the dragging of inertial frames \cite{Zimmerman:kv}.
In fact one can interpret the Schwarzchild-NUT spacetime 
as consisting of two counter-rotating regions, with a vanishing total angular momentum~\cite{Manko:2005nm,Kleihaus:2013yaa}.

Our results show the following main trends. Firstly, for the EGBs model, but not for the ECSs case, scalarised solutions exist for the vanishing NUT charge case, $n=0$. This is easy to understand from the fact that only for $n\neq 0$, the Pontryagin density is excited in the scalar-free case, $cf.$ eq.~(\ref{LCS0}). This basic difference leads to a distinct domain of existence of solutions in the CS case, in particular exhibiting a double branch structure. In both cases the domain of existence of scalarised solution, for fixed NUT charge, is limited to a finite interval of coupling parameters $\alpha$. At the lowest coupling, the limiting solutions define the existence line, $i.e.$ the line of scalar-free solutions that can support a scalar cloud, as a test field configuration. At the largest coupling, the limiting solutions define a critical line of maximally scalarised solutions. This interval of couplings wherein scalarised solutions occur varies with the NUT charge as follows. In the GB case, increasing the NUT charge (for fixed $M$) demands a stronger non-minimal coupling for scalarisation; in the CS case, due to the double branch structure, both this and the opposite trend are found. We expect the same trends to occur, in terms of the horizon area rather than NUT charge, in the Kerr case, which is confirmed by preliminary results.

\bigskip

%%%%%%%%%%%%%%%%%%%%%%%%%%%%%%%%%%%%%%%%%%%%%%%%%%%%%%%%%
\begin{figure}[ht!]
\begin{center}
%\subfigure[Mass]
{ \includegraphics[width=8cm]{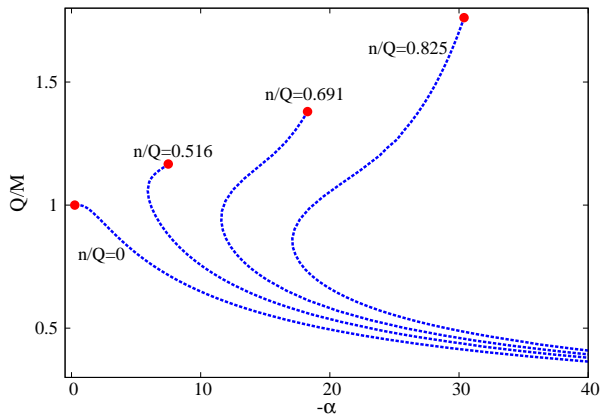}}
\caption{ $k=0$ existence lines for the Reissner-Nordstr\"om-NUT solution for different values of the NUT charge normalised to the electric charge $Q$.   
\label{existence-line-sEMs}
}
\end{center}
\end{figure} 
%%%%%%%%%%%%%%%%%%%%%%%%%%%%%%%%%%%%%%%%%%%%%%%%%%%%%%%%%

Albeit a rather unusual spacetime, the Schwarzschild-NUT solution has been generalised in various directions. In particular, gauge fields have been added~\cite{Brill,Radu:2002hf,Brihaye:2005ak}.\footnote{Another generalisation, within a low-energy effective field theory from string theory is found in~\cite{Johnson:1994ek}. All these solutions share the same properties of the vacuum NUT metric, in particular the same asymptotic and causal structure.} The simplest of these solutions~\cite{Brill} represents an electrically charged version of the Schwarzschild-NUT solution, the Reissner-Nordstr\"om-NUT solution. This spacetime can be scalarised in a different way: by a non-minimal coupling of the scalar field to the Maxwell invariant,~following the recent work~\cite{Herdeiro:2018wub}. That is, we 
take the action (\ref{generalaction}) but with
 \begin{eqnarray} 
{\cal I}=F_{ab}F^{ab} \ ,
\end{eqnarray}
where $F=dA$.
The Reissner-Nordstr\"om-NUT solution can be expressed in the generic form (\ref{metric}),
	with (we consider the electric version of the solution only):
\begin{eqnarray} 
\label{brill}
g=r^2+n^2\ , \qquad \sigma(r)=1\ , \qquad N(r)=\frac{r^2-2Mr-n^2+Q^2}{r^2+n^2} \ ,
\end{eqnarray}
and a $U(1)$ potential
\begin{eqnarray}  
A=\frac{Qr}{r^2+n^2}(dt+2n \cos \theta d\varphi) \ .
\end{eqnarray}
Here $Q$ is interpreted as an electric charge. Choosing a coupling function~\cite{Herdeiro:2018wub}
\begin{eqnarray} 
f(\phi)=e^{-\alpha \phi^2},
\end{eqnarray}
{we have performed a preliminary study of the corresponding scalar clouds.  
The existence lines are shown in Figure~\ref{existence-line-sEMs}
for several values of the ratio $n/M$.
The red dots  at the end of the curves indicate limiting 
configurations
where the function $N(r)$ in (\ref{brill})
develops a double zero
($i.e.$ the extremal Reissner-Nordstr\"om BH for $n=0$).
The new feature for $n\neq 0$
is the existence,
for the same value of the coupling constant,
 of two different unstable Reissner-Nordstr\"om-NUT configurations,
distinguished by the value of the ratio  $Q/M$.

Finally, we mention two other possible directions to bring together scalarisation and the Schwarzschild-NUT solution. Firstly, one could study the properties of the Euclideanised Schwarzschild-NUT solution, within the model (\ref{generalaction}).  Could there be scalarised instantons? In this context, we recall that the  Euclideanised Schwarzschild-NUT solution has found interesting applications in the context of quantum gravity~\cite{Hawking:ig}. Secondly, continuing the Schwarzschild-NUT metric through its  horizon at $r=r_h$ leads to the Taub universe, a homogeneous, non-isotropic cosmology with an  $S^3$ spatial topology. 
 Whereas the  Schwarzschild solution has a curvature singularity at $r = 0$,  this is not the case for $n\neq 0$ and  
the radial coordinate in the Schwarzschild-NUT solution may span the whole real axis.
 Thus,  following  \cite{Brihaye:2016lsx,Brihaye:2016vkv},
it could be interesting
to investigate  the behaviour of the scalarised solutions inside the horizon.

%%%%%%%%%%%%%%%%%%%%%%%%%%%%%%%%%%%%%%%%%%%%%%%%%%%%%%%%%%%%%%%%%%%%%
\section*{Acknowledgements}
%\noindent{\bf{\em Acknowledgements.}}
%%%%%%%%%%%%%%%%%%%%%%%%%%%%%%%%%%%%%%%%%%%%%%%%%%%%%%%%%%%%%%%%%%%%%
This work has been supported by the FCT (Portugal) IF programme, by the FCT grant PTDC/FIS-OUT/28407/2017, by  CIDMA (FCT) strategic project UID/MAT/04106/2013, by CENTRA (FCT) strategic project UID/FIS/00099/2013 and
by  the  European  Union's  Horizon  2020  research  and  innovation  (RISE) programmes H2020-MSCA-RISE-2015
Grant No.~StronGrHEP-690904 and H2020-MSCA-RISE-2017 Grant No.~FunFiCO-777740. The authors would like to acknowledge
networking support by the
COST Action CA16104.

%%%%%%%%%%%%%%%%%%%%%%%%%%%%%%%%%%%%%%%%%%%%%%%%%%%%%%%%%%%%

%%%%%%%%%%%%%%%%%%%%%%%%%%%%%%%%%%%%%%%%%%%%%%%%%%%%%%%%%%%%

 \end{document}